\documentclass[preprints,accept]{Definitions/mdpi}
\firstpage{1} 
\pubvolume{1}
\issuenum{1}
\articlenumber{0}
\pubyear{2024}
\copyrightyear{2024}
\externaleditor{Academic Editor: Firstname \linebreak  Lastname}
\datereceived{6 March 2024} 
\daterevised{2 May 2024} 
\dateaccepted{11 May 2024} 
\datepublished{ } 
\hreflink{https://doi.org/} 

\Title{BRDF-Based Photometric Modeling of LEO Constellation Satellite from Massive Observations}

\TitleCitation{BRDF-Based Photometric Modeling of LEO Constellation Satellite from Massive Observations}

\Author{{Yao Lu} 
 $^{1,2}$\orcidA{}}

\AuthorNames{Yao Lu}

\AuthorCitation{{Lu,} 
Y.}

\address{%
$^{1}$ \quad Purple Mountain Observatory, Chinese Academy of Sciences, Nanjing 210023, China; luyao@pmo.ac.cn\\
$^{2}$ \quad Key Laboratory for Space Objects and Debris Observation, Chinese Academy of Sciences,
\linebreak  Nanjing 210023, China}

\abstract{Modeling the brightness of satellites in large Low-Earth Orbit (LEO) constellations can not only assist the astronomical community in assessing the impact of reflected light from satellites, optimizing observing schedules and guiding data processing, but also motivate satellite operators to improve their satellite designs, thus facilitating cooperation and consensus among different stakeholders. This work presents a photometric model of the Starlink satellites based on the Bidirectional Reflectance Distribution Function (BRDF) using millions of photometric observations. To enhance model accuracy and computational efficiency, data filtering and reduction are employed, and chassis blocking on the solar array and the earthshine effect are taken into account. The assumptions of the model are also validated by showing that the satellite attitude is as expected, the solar array is nearly perpendicular to the chassis, and both the solar array pseudo-specular reflection and the chassis earthshine should be included in the model. The reflectance characteristics of the satellites and the apparent magnitude distributions over station are finally discussed based on the photometric predictions from the model. In addition to assessing the light pollution and guiding the development of response measures, accurate photometric models of satellites can also play an important role in areas such as space situational awareness.}


\begin{document}


\section{Introduction}

LEO Internet constellations, such as Starlink, are being rapidly deployed and will contribute significantly to global network coverage, especially in remote areas, emergency communications and mobile broadband, etc. However, there are growing concerns about its potential negative impacts \cite{Barentine,Green,McDowell}. For instance, having tens of thousands of satellites in orbit could generate additional space debris and heighten the risk of collisions between orbiting objects. This could also potentially impact astronomical observations, both in terms of reflected light from satellites on optical and near-infrared (NIR) \mbox{observations \cite{Bassa,Boley,Hainaut,Hall,Mróz}} and in terms of broadcast signals from satellites on radio observations \cite{dqs1,dqs2}. The astronomical community has recognized the issue of light pollution, especially since the launch of the first Starlink satellites into orbit. Several workshops have discussed feasible solutions to reduce the impact of low-orbit constellations on astronomical observations \cite{dqs1,dqs2}. Astronomers are advised to conduct full-element simulations of satellite streaks on astronomical \mbox{images \cite{SATCON1,SATCON2}}. This will help quantify the impact of reflected light from satellites on astronomical observations and the loss of scientific objectives. The aim is to assist the astronomical community in better understanding and addressing the challenges of satellite constellations on astronomy, and to take appropriate measures to mitigate their impact. Astronomers are also encouraged to conduct satellite photometric observations under a wider range of observation geometries in order to characterize the complex reflective features of satellites \cite{SATCON1,SATCON2}, including both slowly varying and fast-varying features such as flashes. This work will serve as a foundation for bright satellite avoidance in future astronomical scheduling \cite{Hu}. It will also allow testing of the effectiveness of satellite brightness mitigation by satellite operators through comparisons of the brightness distributions of different satellite versions. The simulation of satellite streaks and the characterization of satellite brightness require the establishment of satellite photometric models. These models must accurately predict satellite brightness at varying solar altitudes and observation directions.

BRDF is  commonly used to describe the optical reflectance behavior of satellites with complex shapes. The modeling of space object brightness and its temporal variation using BRDF have been widely used to study parameter inversion problems related to space object brightness. e.g., the estimation of the rotational motion parameters of rocket bodies, the determination of the rotational state of defunct satellites, and the speculation of the surface materials of satellites. For the LEO constellations that have been deployed, some photometric modeling studies have used simple BRDF models, such as the diffuse spherical \mbox{model \cite{Lawler,Osborn}}, polynomial model \cite{Mallama1,Mallama2}, Minnaert model \cite{Horiuchi1,Tregloan}, etc. While these models can partially capture the characteristics of the spatial distribution of the reflected light, they cannot obviously compare with the accuracy of another class of true shape-based BRDF models. Such a BRDF model of the Starlink visor-sat by decomposing the satellite into the chassis and solar array was developed accounting for the blocking effects of its different parts \cite{Cole}. However, the model does not consider earthshine contributions, which may result in reduced fidelity in the observation geometry dominated by earthshine. The BRDF models of the Starlink V1.5 satellite were established based on laboratory measurements and astronomical observations \cite{Fankhauser}. The observational best-fit model also includes reflections from the chassis and solar array, and accounts for the earthshine effects of the chassis in the model. However, the model fitting does not include the earthshine component due to insufficient coverage of the earthshine-dominated geometry of the data. Additionally, the model does not take into account the blocking of the solar array by the chassis and the backward scattering of the solar array (corresponding to the lampshade effect of the sail mentioned in the statement from SpaceX \cite{spacex}). Improving the accuracy of satellite photometric models is urgently needed for issues such as the quantitative assessment of the impact of satellite streaks, verification of the mitigating effect of satellite brightness, development of overbright satellite avoidance algorithms and application to space situational \linebreak  awareness, etc.

Previous studies on photometric modeling of Starlink satellites have established a strong foundation for achieving more accurate and higher spatial resolution photometric models, which is the primary objective of this work. In this paper, we utilize {millions of observations} with more comprehensive coverage of the observation geometry to fit the model. Data filtering and data reduction steps are employed to enhance model accuracy and speed up calculations. The model considers both shadowing and earthshine effects to improve satellite photometric prediction. Additionally, the satellite attitude assumptions, the solar array opening angle assumptions, and the necessity of earthshine are also examined during the fitting process  to ensure the final model is sound. Based on the number of LEO constellation satellites launched into orbit by January 2024, as shown in Table~\ref{tab1}, this paper selects the Starlink V1.5 satellite, which has the largest number of satellites, as the modeling target. 

{Section} 
 \ref{Section 2} outlines the methodology for building the BRDF-based photometric model. {Section} \ref{Section 3} describes the photometric data used for model fitting and its pre-processing procedures. {Section} \ref{Section 4} includes several details of the model-fitting process. {Section} \ref{Section 5} presents the photometric prediction results of the model, and the shortcomings of our model, as well as possible improvements are finally discussed.

\begin{table}[H] 
\caption{Number of LEO constellation satellites launched into orbit. Only currently deployed constellations that number more than one hundred, i.e., Starlink and Oneweb, are considered. Satellite counts from Jonathan's space report (\url{https://www.planet4589.org}, {accessed on 12 January 2024}
) and Wikipedia (\url{https://en.wikipedia.org/wiki/List\_of\_Starlink\_and\_Starshield\_launches}, {accessed on 12 January 2024}).\label{tab1}}
\newcolumntype{C}{>{\centering\arraybackslash}X}
\begin{tabularx}{\textwidth}{CCCcC}
\toprule
\textbf{Satellite}	& \textbf{Starlink v1.0}	& \textbf{Starlink v1.5}& \textbf{Starlink v2.0mini}& \textbf{Oneweb}\\
\midrule
\textbf{Count} & 1665  &  2987& 978  & 640\\
\bottomrule
\end{tabularx}
\end{table}

\section{Photometric Model \label{Section 2}}

A simplified satellite shape model is first constructed based on publicly available information. The reflection characteristics of each plane are then described using the Phong BRDF \cite{Phong}. To fit the model, the geometric variables from various observations need to be transformed to the satellite body-fixed coordinate system, which is calculated based on the satellite's two-line element (TLE) data and the assumed attitudes. The observation in satellite on-station phase is the only data used for fitting. This is because the attitude of the satellite is known at this stage. Additionally, the integrated satellite BRDF model must be corrected for the blocking effect of the chassis on the incident and reflected light from the solar array. The modeling of the external light source considers both direct sunlight and sunlight reflected from the Earth's surface (earthshine) \cite{Horiuchi1,Fankhauser}. Finally, the photometric prediction model is established by converting the flux given by the BRDF model to astronomical magnitude based on the apparent magnitude of the Sun.

\subsection{Satellite BRDF Model}{\label{brdf}}

{Following} the method used in \cite{Cole}, we disassemble the main body of Starlink satellite {into two panels: the chassis and the solar array}. Diffuse and specular reflections are considered for each surface by using the Phong BRDF \cite{Phong}. In fact, due to the limitations of the satellite's orbital altitude (e.g., 550 km) and attitude during normal operation, the standard specular reflection of the chassis cannot be observed on Earth, but the sidelobe close to the specular direction can be observed, making the satellite significantly brighter. The pseudo-specular reflection here refers mainly to the specular sidelobe, also called the glossy component. The exponent of the cosine function in the Phong BRDF is used to model the intensity distribution of the specular sidelobe. The larger the exponent argument, the more concentrated the specular energy, and the narrower the sidelobe footprint. For diffuse reflection, the Lambertian model is used.

Lambertian reflection is also used to model the back-scattering from the solar array, to explain the contributions of light leakage from the gaps and edges of the array to the brightness \cite{Cole,spacex}. Our fitting results seem to support the addition of this term, otherwise the brighter region at the back of the solar array cannot be better explained. However, another convincing explanation that this brightness excess comes from earthshine reflected from {the chassis} can also make remarkable contributions in the backward direction of the solar array. As a result, although both the backward scattering from the solar array and the reflected earthshine light from the chassis are considered in our model, the coupling of the two to some extent makes it {difficult to identify the specific contributions of each component}.

The calculation of the reflection for each plane in our model is discussed in the body-fixed reference frame, which is defined by the chassis center and the chassis panel. The unfolded solar array is considered perpendicular to the chassis in most references, but Cole believes that it may deviate from the vertical direction with a constant panel offset angle (POA) \cite{Cole}. As the most naive assumption, it seems that an opening angle that changes with the solar altitude {is the appropriate variable to use when taking} power generation efficiency into account. In order to test which hypothesis can better explain the {observational} data, the opening angle is set as a free parameter in our model and fit at different solar altitudes. See Section{~\ref{sail_vertical}} for a detailed discussion of the angle.

In summary, the integrated satellite BRDF $B_{total}$ (includes the cosine function and area) is derived from five parts, {i.e.,} the specular and diffuse reflections of the solar array, the backward scattering of the solar array, and the specular and diffuse reflections of chassis: 
\begin{linenomath}
\begin{align}
B_{total} &= B_{s,sail} + B_{d,sail} + B_{b,sail} + B_{s,base} + B_{d,base} \\
B_{s,sail} &= c_1\frac{n_{sail}+2}{2\pi}\cos^{n_{sail}}\theta_{sail}\cos\phi_{i,sail}\cos\phi_{r,sail}A_{sail}(1-BF) \\
B_{d,sail} &= c_2\frac{1}{\pi}\cos\phi_{i,sail}\cos\phi_{r,sail}A_{sail}(1-BF)\\
B_{b,sail} &= c_3\frac{1}{\pi}\cos\phi_{i,sail}\cos(\pi-\phi_{r,sail})A_{sail}(1-BF')\\
B_{s,base} &= c_4\frac{n_{base}+2}{2\pi}\cos^{n_{base}}\theta_{base}\cos\phi_{i,base}\cos\phi_{r,base}A_{base}\\
B_{d,base} &= c_5\frac{1}{\pi}\cos\phi_{i,base}\cos\phi_{r,base}A_{base}
\end{align}
\end{linenomath}
where $c_1,c_2,c_3,c_4,c_5,n_{sail},n_{base}$ are the basic coefficients of our BRDF model to be fit. $\theta$ is the angle between the directions of the observer and the specularly reflected beam for the solar array or chassis. $A$ is the area of the solar array or chassis plane. $\phi_i, \phi_r$ are the angles of the incident and reflected directions relative to the normal direction of the plane. $BF$ is the blocking factor, and $BF'$ is the special case that only incident light blocking is accounted for. Their calculations will be described below. Note that the light source for the chassis includes additional earthshine reflection in our model, which means $B_{s,base},B_{d,base}$ will be the sum of \linebreak  multiple components.  

\subsection{Calculations of Geometric Arguments}

Because only on-station phase data with known satellite attitude will be used to fit the model, the valid time period for the on-station phase is first identified. Considering that these satellites have been in almost circular orbits, the orbital semi-major axes are used to determine their on-station time range. In detail, the historical two-line element (TLE) records of observed Starlink satellites are retrieved from the Space-Track.org website, by which one can calculate the changes of the semi-major axes over time. We assumed the satellite is on-station only when the semi-major axis {is} 530 km greater than Earth's radius (actually, the height of their operational orbits had slight variances with 550 km; 20 km is a safe threshold); meanwhile, the variation is less than 100 m/day over three adjacent TLE records. In this way, we obtained the period of the on-station phase for each observed {satellite}. According to the information from SpaceX \cite{spacex}, when Starlink arrives at operational altitude, its configuration will be transformed to a shark-fin mode, basically consisting of a flat chassis facing the nadir and a vertical solar array facing the Sun. 

To model the brightness of the satellite with BRDF, a body-fixed reference frame is set up, where the origin is the center of the chassis, axis $z$ is the normal of the chassis plane pointing to the local zenith, axis $x$ lays in the chassis plane and always makes the direction of the Sun be in the $zx$ plane ($\alpha_{sun}\equiv0$), and axis $y$ complements the other two axes to form a right-handed coordinate system. Now, we can derive the coordinates of the Sun and the observer in the body-fixed frame, through their coordinates and the satellite coordinates in the Geocentric Celestial Reference System (GCRS). Using a spherical coordinate, the direction of the Sun in the body-fixed frame is $(0, \delta_{sun})$, the observer $(\alpha_{obs}, \delta_{obs})$, where $\alpha$ and $\delta$ represent the longitude and latitude angle, respectively, and are derived as follows. 

First, the closest TLE record in time for each observation is retrieved to calculate the position of satellite $R_{sat}$ at the observational epoch in the GCRS using the \textit{Skyfield} package \cite{Skyfield}. The positions of the Sun and observatory $R_{sun}, R_{obs}$ at the same epoch in the GCRS are obtained by the \textit{Astropy} package \cite{astropy}, and the positions relative to the satellite $r_{sun}, r_{obs}$ are
\begin{linenomath}
\begin{align}
r_{sun} &= R_{sun} - R_{sat}\\
r_{obs} &= R_{obs} - R_{sat}
\end{align}
\end{linenomath}

The zenith direction unit vector relative to the satellite is:
\begin{linenomath}
\begin{equation}
r_{z} = \frac{R_{sat}}{|R_{sat}|}
\end{equation}
\end{linenomath}

Now, we can derive $\delta_{sun},\delta_{obs},\alpha_{obs}$ as follows:
\begin{linenomath}
\begin{align}
\delta_{sun} &= \frac{\pi}{2} - Ang(r_{z},r_{sun})\\
\delta_{obs} &= \frac{\pi}{2} - Ang(r_{z},r_{obs})\\
\alpha_{obs} &= \left\{ \begin{array}{ll}
Ang(r_{z}\times r_{sun}, r_{z} \times r_{obs}) & \mbox{if $(r_{sun} \times r_{obs})\cdot r_{z} \ge 0$}\\
2\pi - Ang(r_{z} \times r_{sun}, r_{z}\times r_{obs}) & \mbox{if $(r_{sun} \times r_{obs})\cdot r_{z} < 0$}  \\
                   \end{array}\right.
\end{align}
\end{linenomath}
Here, $Ang(v_{1},v_{2})$ denotes the angle between two vectors within the range of $[0,\pi]$:

\begin{linenomath}
\begin{equation}
Ang(v_{1},v_{2}) = arccos(\frac{v_1\cdot v_2}{|v_1||v_2|})
\end{equation}
\end{linenomath}

Given $\delta_{sun},\delta_{obs},\alpha_{obs}$ in the satellite body-fixed frame, it is easy to derive the angles of the incident or reflected light directions $\phi$, as well as the offset angles of the specularly reflected beams $\theta$ for each satellite element.

The intensity of sunlight at satellites is assumed to be constant during the observation period. The distance from the observatory to the satellite is used to correct the original magnitude to the normalized magnitude at the range of 1000 km by:
\begin{linenomath}
\begin{equation}
m_{nor} = m_{ori} - 2.5\log(\frac{|r_{obs}|}{1000})^2
\end{equation}
\end{linenomath}
where $|r_{obs}|$ is the range in km and $m_{ori}$ and $m_{nor}$ are the original and normalized magnitude of the satellite. By the range correction, these three parameters $\delta_{sun}, \alpha_{obs}, \delta_{obs}$ are all geometrical conditions required by the BRDF model.

\subsection{Blocking Effect}

{\textls[-25]{The partial blocking of the solar array by the chassis} is accounted for in our model. According to the working attitude of the Starlink satellite, the solar and viewing directions corresponding to the observed moments are always below the chassis in the body-fixed frame, making the solar array obscured from both illumination and visibility. The invisible or unilluminated fraction of the solar array, called the blocking factor here, $BF(\delta_{sun},~\alpha_{obs},~\delta_{obs},~POA,~l_{sail},~w_{sail},~l_{base},~w_{base})$} can be calculated by basic graphical projection techniques. As the angle between the solar array and chassis in our model is allowed to be off vertical, the POA is included in the arguments. Here, $l_{sail},~w_{sail},~l_{base},~w_{base}$ are the length and width of the solar panel and chassis separately, set to $8.1,~2.8,~1.3,~2.8$ m in our model.

During the satellite on-station phase, the Sun is always located directly in front of and below the chassis, making the unilluminated area of the sail a rectangle at its bottom. The ratio of this area $BF'$ can be represented as:
\begin{linenomath}
\begin{equation}
BF' = \frac{l_{base}}{l_{sail}}\frac{-\sin{\delta_{sun}}}{\cos(POA-\delta_{sun})}
\end{equation}
\end{linenomath}

\begin{figure}[H]
\includegraphics[width=10 cm]{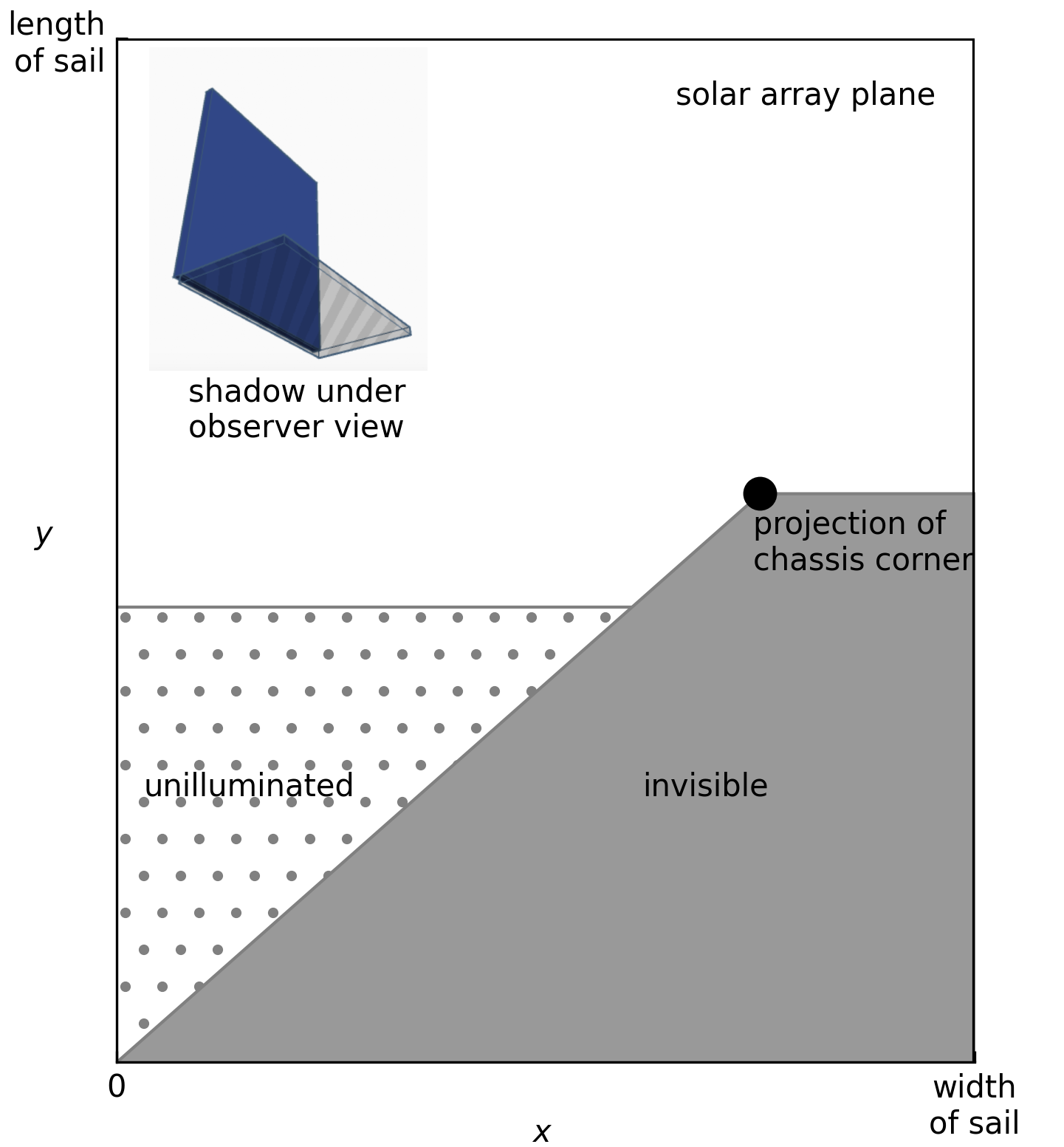}
\caption{{The} 
 sketch of chassis blocking solar array. The blocking factor is the percentage of unilluminated and invisible area to the whole sail area. The calculation of the coordinates of the projection point is shown in Equations \eqref{eq16}--\eqref{eq18}.\label{fig1}}
\end{figure}

The blocking area in the viewing direction is more difficult to determine, because of the additional observation azimuth factor. Considering the symmetry of the satellite model with respect to the viewing azimuth, the projection point of the chassis corner along the viewing direction on the sail can be determined as follows: 
\begin{align}
\label{eq16}
x_p &= l_{base} l \cos{\delta_{obs}} \sin{\alpha_{obs}}
\end{align}
\begin{align}
\label{eq17}
y_p &= l_{base} \sqrt{(1-l\cos{\delta_{obs}}\cos{\alpha_{obs}})^2 + l^2\sin^2\delta_{sun}}
\end{align}
\vspace{-16pt}  
\begin{align}
\label{eq18}
l &= \frac{\cos{POA}}{\cos\delta_{obs}\cos\alpha_{obs}\cos{POA}+\sin\delta_{obs}\sin{POA}}
\end{align}
where $(x_p,y_p)$ is the projection point on the solar array, as shown in Figure~\ref{fig1}. Then, the ratio of the invisible or unilluminated area of the sail to the whole area of the sail can be easily calculated (omitting the calculation process here) based on the location of this projection point.

\subsection{Earthshine Effect}

For LEO satellites, in addition to direct solar irradiation, reflected sunlight from the Earth's surface sometimes can also act on the surface of the satellite, providing an additional radiative pressure, on the one hand, which has been widely studied in the field of precise orbit determination, and on the other hand, increasing the reflection of the satellite, but with a much smaller intensity that is not widely considered in satellite photometric modeling. For modeling the full-angle brightness of Starlink satellites, especially in the observational directions where earthshine is significant, it was considered by \cite{Horiuchi1,Horiuchi2}, and a discretized numerical calculation of earthshine was described in detail in \cite{Fankhauser}. In this paper, we also use a discretized approach, by first meshing the Earth's surface by the \textit{Healpix} tool, with the degree of thinness determined by the model accuracy required for earthshine. Note that this discretized numerical approach is the most time-consuming step in the overall model fitting and, thus, requires a trade-off between model accuracy and computational volume. Then, given the positions of the satellite, the Sun, and the observatory in the GCRS, the sunlight reflected by a single grid element ({simplified to be flat}) and further diffusely and specularly reflected on the satellite's chassis can be computed through a method similar to the geometric calculations described above. Here, the reflective properties of the Earth's surface are simplified to Lambertian reflections, whose albedo is set to 0.3. The diffuse and specular reflection coefficients of the chassis are set to be consistent with directly reflected sunlight. Finally, the results of all grids that can reflect to the chassis are summed up to obtain the final reflected flux of earthshine.

\subsection{Flux to Magnitude}

The flux density of satellite-reflected light in the Johnson-V band at the observer is: 
\begin{linenomath}
\begin{align}
f_{v,sat} &= \frac{f_{v,sun}B_{total}}{d^2}\\
f_{v,sun} &= 10^{\frac{m_{v,sun}-m_{v,0}}{-2.5}}
\end{align}
\end{linenomath}
where $f_{v,sun}$ is the solar Johnson-V band flux density in the vicinity of the Earth, $m_{v,sun}$ is the solar apparent magnitude in the Johnson-V band and set to a constant value of $-26.76$ mag here \cite{Willmer}, ignoring variations in the Sun--Earth distance and the intensity of the solar radiation, $m_{v,0}$ is the zero-point in the band, and $d=1,000$ km is the normalized range. Thus the, satellite reflected light magnitude {is:} 
\begin{linenomath}
\begin{align}
m_{v,sat} = -2.5\log{f_{v,sat}}+m_{v,0} = -2.5\log\frac{B_{total}}{d^2}+m_{v,sun}
\end{align}
\end{linenomath}

\section{Data for Model Fitting \label{Section 3}}

The Mini-MegaTORTORA (MMT-9) public database \cite{Karpov,Beskin} is employed to fit our model, which is also used in the studies by \cite{Mallama1,Mallama2,Cole}. It is used primarily for large field-of-view time-domain survey, but also records the brightness of passing space objects. The magnitudes are reported within 0.1 mag errors of the V-band \cite{Mallama1}. As of {January} 
 2024, it has recorded more than {four million data points} of Starlink V1.5 satellites. The voluminous dataset is advantageous for model fitting due to its extensive sky coverage. However, it requires meticulous preprocessing to account for possible issues such as target association errors and systematic deviations between instruments. To obtain the suitable data for fitting, we executed a preprocessing step followed by data reduction.

The magnitudes extracted from the data will be filtered as follows: First, exclude magnitudes in all other bands except for the clear filter. Next, remove observations of satellites located in the penumbra using a simple criterion to determine the penumbra range, i.e., the height of the connection line between the Sun and satellite being lower than the 100 km Kármán line. Eliminate data with incorrect associations that have impossible visibility for the MMT-9 station. Finally, standardize all magnitudes to a range of 1000 km.
                
There are still millions of magnitudes after filtering. Given a solar altitude range like a 1 degree width, for a {patch-like} angular distance within 1 degree around a direction in the body-fixed frame, there are typically hundreds of magnitudes from different satellites at various times. Because the {model parameters} that determine satellite brightness for these {data points are} nearly identical, the magnitudes within a small patch should be very similar for each satellite version. However, even on a smaller scale where the spatial variation of the model is negligible, magnitudes still exhibit up to 1 mag $(1\sigma)$ of dispersion. The potential causes include not only observation noise and incorrect target association, but also coordinate transformation errors resulting from satellite deviation from the assumed attitude. For example, to maintain orbital altitude, adjustment of the satellite attitude may be necessary to control the propulsion direction. The Starlink ephemeris indicates regular maintenance occurs several times a week, lasting approximately five minutes each time. The causes of attitude deviation are diverse, yet during most of the on-station phase, the satellite should maintain the assumed working attitude with the chassis pointing to nadir and the solar panel facing the Sun. This means most of the observations within a patch should present almost identical magnitudes. {Taking this into account,} the following method is used to extract effective data for model fitting. The magnitudes are divided into layers based on solar altitude in the body-fixed frame. The spherical surfaces composed of the observer's azimuth and altitude of each layer is segmented using \textit{Healpix} \cite{HEALPix}. Then, the magnitude with the maximum probability density of all magnitudes within each pixel is extracted. Usually it is very close to the median of magnitudes in the pixel. The extracted magnitudes will represent pixel data for model fitting. As a result, millions of magnitudes are compressed to {around twenty thousand magnitudes}. 

The degree of data dispersion within a pixel reflects the error level of the extracted magnitude. We used the interquartile range (IQR) of the pixel data to represent this error, which is typically less than 1 mag and is involved in the fitting as the data uncertainty. 

\begin{figure}[H]
\includegraphics[width=13.8 cm]{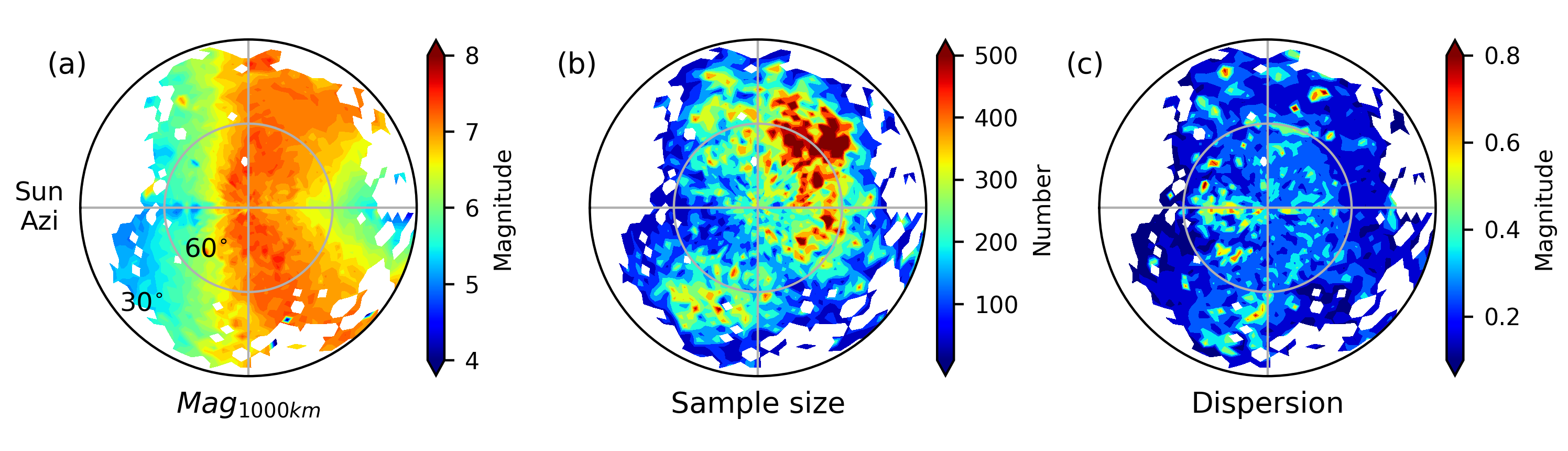}
\caption{Slice of the distribution of the extracted photometric data. In the satellite body-fixed coordinate system, filtered observations are shown here for the solar altitude within $-19^{\circ}$$\sim$$-18^{\circ} $, and the solar azimuth is fixed at the position as shown in the figure. The sky is divided in the reflecting directions using nside = 32 in \textit{Healpix}. (\textbf{a}) Extracted median magnitudes, which are normalized to a 1000 km distance. (\textbf{b}) Number of samples in the grid. (\textbf{c}) Magnitude dispersions within the grid.\label{fig2}}
\end{figure}

Figure~\ref{fig2} is the distributions of the extracted magnitude, the sample number, and the dispersion per pixel in the body-fixed frame with $\delta_{sun}\in(-19^{\circ},-18^{\circ})$. By setting the pixel scale as nside = 32 in \textit{Healpix}, the spatial resolution is sufficiently detailed to characterize brightness variations. Figure~\ref{fig2} also shows the high completeness of the sky coverage for ground-based observations. \vspace{-3pt}  

The high symmetry of the photometric distribution with respect to the solar azimuth can also be seen in Figure~\ref{fig2}{a}
, which is determined by the symmetric shape of the satellite, but also proves that the assumption about the satellite attitude, i.e., that the front of the satellite is always facing the Sun, is valid. Therefore, the data in each layer are folded into solar azimuths from 0 to $\pi$, so that the sample size in each pixel is doubled and the number of extracted magnitudes is reduced to half.

\begin{figure}[H]
\includegraphics[width=11 cm]{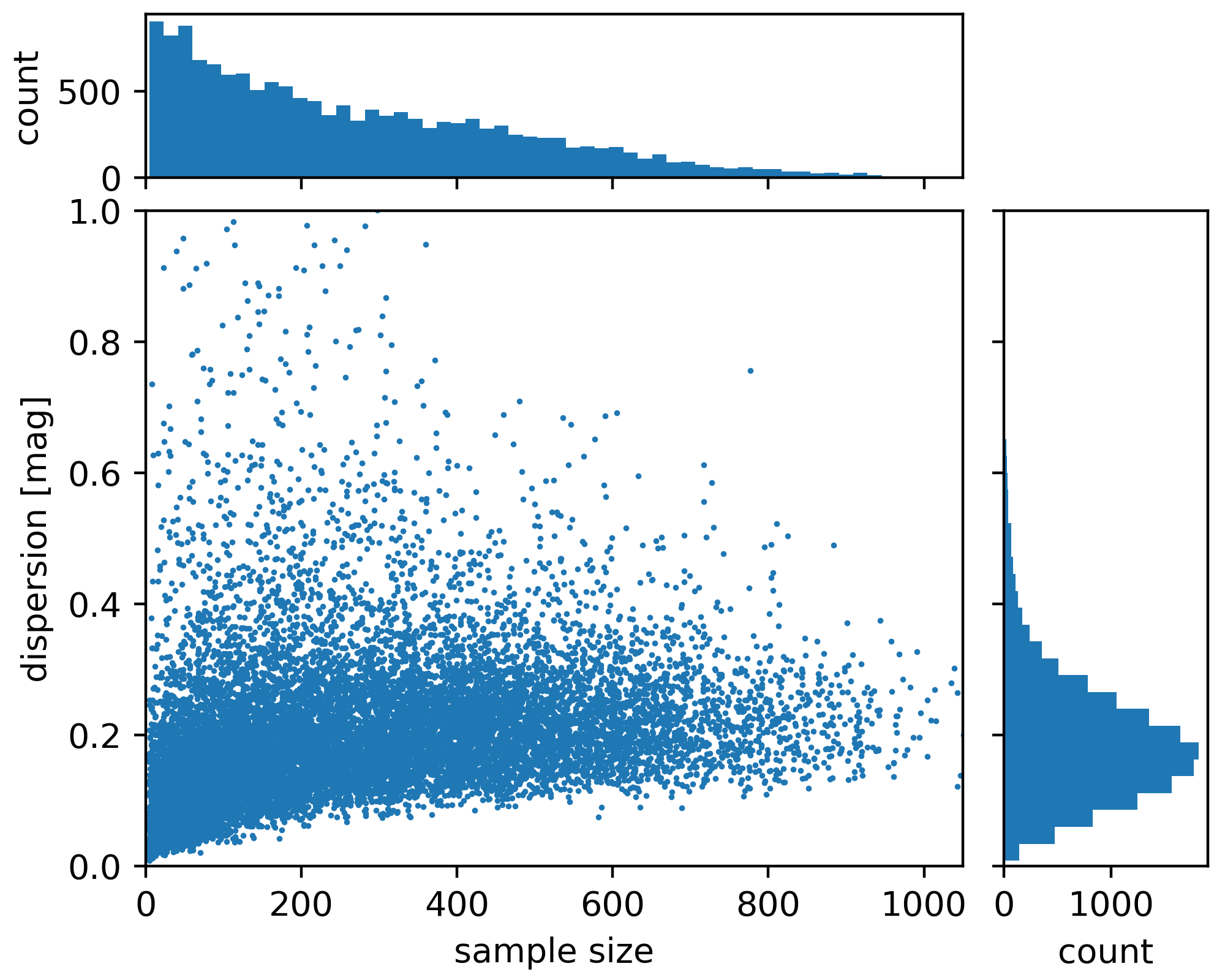}
\caption{{Sample} 
 size vs. dispersion distribution for all grids. The effective grids used for fitting are chosen based on the sample size and dispersion within a grid. After testing, the thresholds of 80 and 0.4 mag are used, by which a better balance between data size and data quality can be achieved.\label{fig3}}
\end{figure}

Only those pixels with large sample sizes and small discretizations will be used. Their cumulative histograms and joint distribution are shown in Figure~\ref{fig3}. A more restrictive threshold improves the quality of the fitting data, but also results in less geometric coverage. After comprehensive tests, the number threshold is set to 80 and the dispersion threshold to 0.4 mag.

Considering that the photometric error of the MMT itself is about 0.1 mag, factors affecting less than 0.1 mag are not corrected in our model, including the effects of variations in the distance between satellites and the Sun, as well as the even smaller influence of cyclic variations in the intensity of solar radiation. 

\section{Model Fitting \label{Section 4}}

Before the final model is fit, possible redundant parameters in the initial model needed to be examined to prevent overfitting and to {simplify the model} as much as possible. To test the assumption that the sail is perpendicular to the chassis, the necessities of the specular reflections on the sail front side and earthshine effect are reported.

\subsection{Verticality of Solar Array}{\label{sail_vertical}}

As discussed above, most of the references assume that the solar array is perpendicular to the chassis; however, Cole adds an extra fixed offset angle to interpret SpaceX's efforts to mitigate the satellite brightness by decreasing the visibility of the sail \cite{Cole}. In fact, adjusting the opening angle should balance sail visibility and power generation efficiency, and it appears that an opening angle that varies with solar altitude in the body-fixed frame makes more sense. Therefore, we also set the offset angle as the fitting parameter in our model, and tested the fitting results at different solar altitudes, respectively.
The data are divided into layers at $1^{\circ}$ intervals by solar altitudes, and those layers with more extensive geometric coverage are used to fit the photometric model, respectively. If the offset angle is correlated with solar altitude, it is hoped to see such a correlation in the fitting results. The results are shown in Table~\ref{tab2}, where the offset angle obtained from the fit changes very little around the vertical, despite the solar altitude changing by $10^{\circ}$. The results hardly suggest that the offset angle varies regularly with solar altitude, but instead verify that it is always almost zero, and therefore, the final photometric model adopts the assumption that the sail is perpendicular to the chassis.

\begin{table}[H] 
\caption{Fitting results for the solar panel offset angle (POA) at different solar altitudes. Other solar altitudes are not considered because the data have a much poorer coverage of the observation geometry for a meaningful fitting. The trend of the POA with solar altitude is not found in the results, and therefore, the model uses a vertical solar panel setup, i.e., the POA is zero.\label{tab2}}
\newcolumntype{C}{>{\centering\arraybackslash}X}
\begin{tabularx}{\textwidth}{CCC}
\toprule
\textbf{Sun Alt \textsuperscript{1}}	& \textbf{POA}	& \textbf{RMSE [mag]}\\
\midrule
$-11^{\circ}$$\sim$$-12^{\circ}$ & $-2.8100^{\circ}$ & $0.2275$ \\
$-12^{\circ}$$\sim$$-13^{\circ}$ & $-2.6224^{\circ}$ & $0.2142$ \\
$-13^{\circ}$$\sim$$-14^{\circ}$ & $-2.4771^{\circ}$ & $0.2245$ \\
$-14^{\circ}$$\sim$$-15^{\circ}$ & $-3.1621^{\circ}$ & $0.1912$ \\
$-15^{\circ}$$\sim$$-16^{\circ}$ & $-2.9903^{\circ}$ & $0.1962$ \\
$-16^{\circ}$$\sim$$-17^{\circ}$ & $-2.3624^{\circ}$ & $0.1886$ \\
$-17^{\circ}$$\sim$$-18^{\circ}$ & $-3.1250^{\circ}$ & $0.2160$ \\
$-18^{\circ}$$\sim$$-19^{\circ}$ & $-3.0606^{\circ}$ & $0.2290$ \\
$-19^{\circ}$$\sim$$-20^{\circ}$ & $-2.3758^{\circ}$ & $0.2230$ \\
$-20^{\circ}$$\sim$$-21^{\circ}$ & $-2.7825^{\circ}$ & $0.2263$ \\
\bottomrule
\end{tabularx}
\noindent{\footnotesize{\textsuperscript{1} In body-fixed frame.}}
\end{table}

\subsection{Specular Reflection of Solar Array}{\label{specular_sail}}

In the on-station phase, both the solar and the viewing directions are located below the chassis. When the assumption of a perpendicular sail is used, for the observations in this paper, the sail specular reflection beam differs from the viewing direction by at least $30^{\circ}$. The specular reflection lobe in the Phong BRDF model can theoretically be scaled up to such a large angle, but when the fitting data reach only the edges of the lobe region, it is not clear whether the fitted model can capture this reflection, in other words, whether it is necessary to include this reflection in the model. To verify this uncertainty, the model is fit with and without specular reflection of the solar array, respectively. More degrees of freedom of the model always lead to smaller fitting root-mean-squared error (rmse), as 0.2408 and 0.2423. The Akaike Information Criterion (AIC) \cite{aic} is used to further compare alternative models. {The model within this reflection is supported because of its smaller {AIC value of $-$13,232} 
 against another {$-$13,176}
. To prove the improvement between the models, a paired \emph{t}-test for the null hypothesis that the fitting residuals of two models have identical average is performed. The \emph{p}-value is very close to 0, which means the performances of the two models are statistically different. Thus, our final model includes this reflection.}

\subsection{Necessity of Earthshine}

The earthshine effect is relatively undiscussed in satellite photometric modeling, partly because it is more difficult to formulate and the parameters related to the Earth's albedo are hard to determine accurately, also because its contribution is relatively smaller and can usually be ignored. {Since the fitting rmse of the models with and without earthshine contribution are very close (0.2408 against 0.2410), a paired \emph{t}-test on their residuals is used to test the necessity of the earthshine effect for our model. The \emph{p}-value is about 0.25, meaning that two sets of the residuals are statistically identical in their means. Therefore, the earthshine reflection is not included in our final model}. As discussed above, the distribution characteristics of the sail back-scattering are similar to those of earthshine from the chassis within a certain range, and thus, it is difficult to distinguish well the coupled contributions from each other in the model fitting. This means that, even if the earthshine effect is removed, a part of its contribution can actually be absorbed by the sail back-scattering component. Considering that both the earthshine effect and sail back-scattering have real and clear physical significance for the reflections of the satellite, our model includes both of them, implying that the coupling of the relevant model parameters is unavoidable.

\subsection{Final Model}

\begin{figure}[H]
\includegraphics[width=13.8 cm]{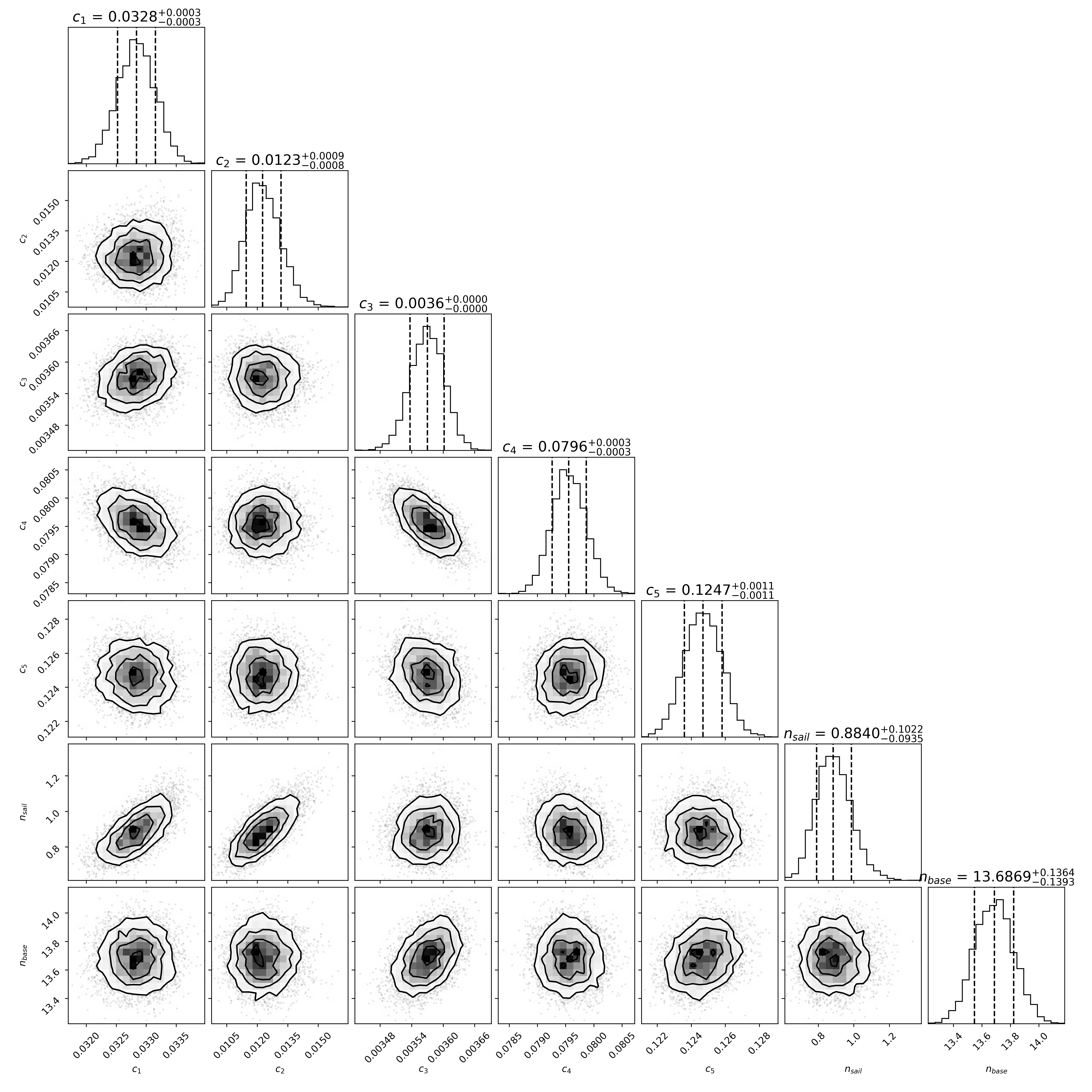}
\caption{{The} 
 posterior distributions of the model parameters from the Markov Chain Monte Carlo (MCMC) sampling \cite{emcee}. {The uninformative priors with boundary constraints are applied. A total of 5,000 steps are taken, and the first 200 steps are burn-in. The dashed lines represent the 0.16, 0.5, and \linebreak  0.84 quantiles of the samples, respectively. According to the results, the errors in the parameter estimates are quite small and there is no significant correlation between the parameters.} \label{fig_mcmc}}
\end{figure}

\begin{figure}[H]
\includegraphics[width=13.8 cm]{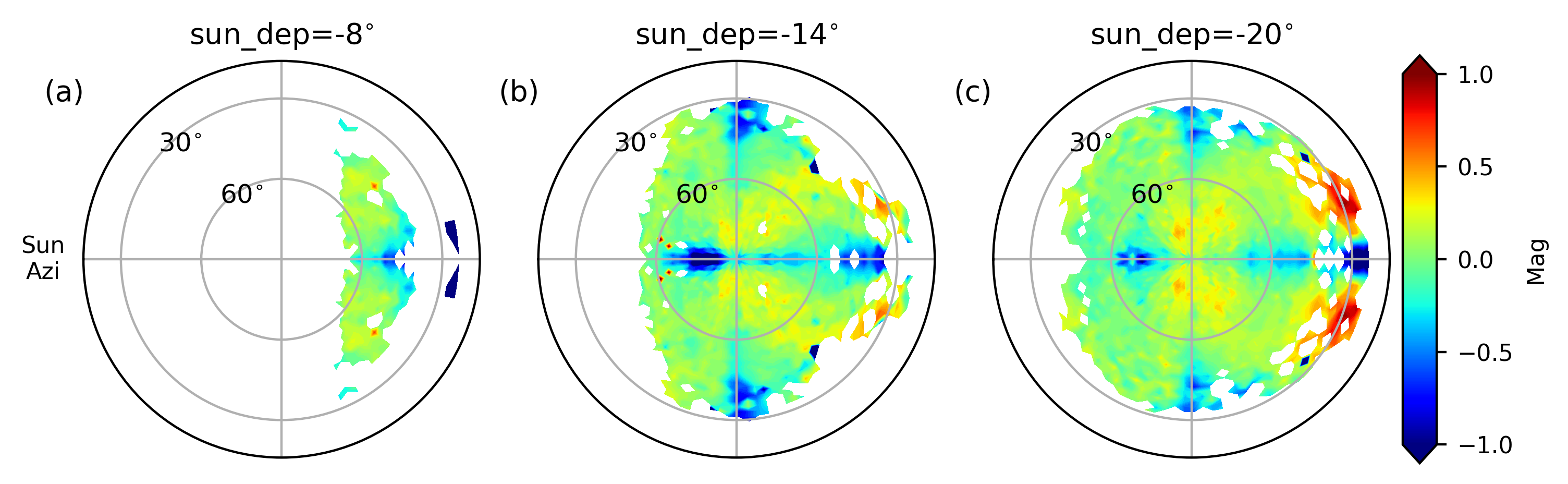}
\caption{{The O$-$C} 
 distributions of model fitting at different solar altitudes. (\textbf{a}) The solar altitude is $-8^{\circ}$. (\textbf{b}) The solar altitude is $-14^{\circ}$. (\textbf{c}) The solar altitude is $-20^{\circ}$. Due to the symmetry of the photometric distribution in the body-fixed frame, the data are folded to $0$$\sim$$180^{\circ}$ along the viewing azimuth before being counted. The effective grid is, therefore, half a circle, with the other half being a mirror. Here, sun\_dep refers to the solar altitude in the body-fixed frame to distinguish from sun\_alt in the Altaz frame.\label{fig4}}
\end{figure}

\begin{figure}[H]
\includegraphics[width=10.8 cm]{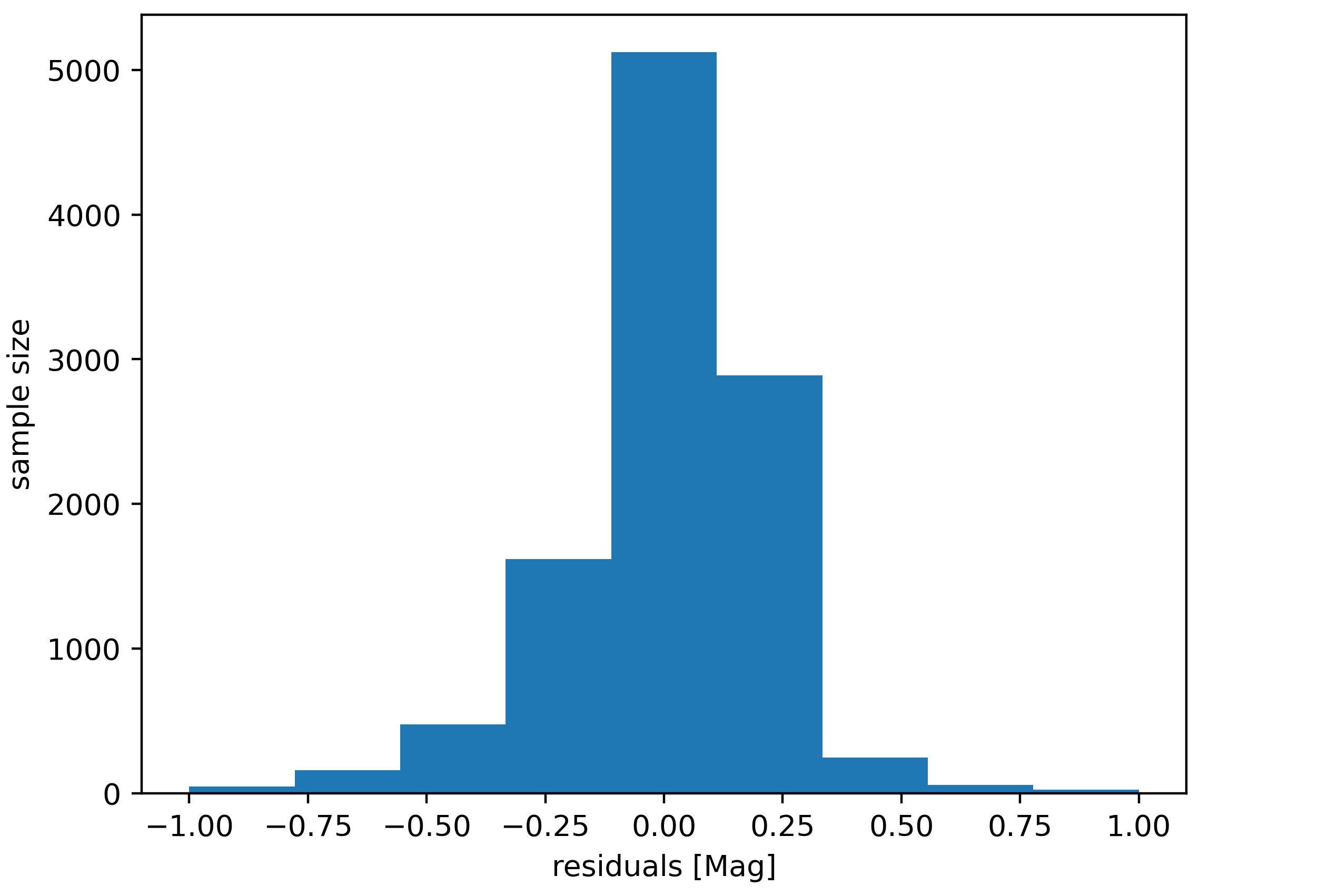}
\caption{Histograms of the fitting residual O-C at all solar altitudes. Given the comprehensive coverage of the observation geometry of the fitting data, the forecast accuracy ($1\sigma$) of the model can be considered comparable to that of the residual rmse of around 0.25 mag.\label{fig5}}
\end{figure}

The final photometric model of the Starlink V1.5 satellite adopts the assumption that the sail is vertical, accounting for the specular reflection of the sail and the earthshine effect. {The posterior distributions of the model parameters are shown in }Figure~\ref{fig_mcmc}, and the mathematical formulas and the meanings of the model parameters can be found in Section~\ref{brdf}. The fitting residuals at different solar altitudes in the body-fixed frame are shown in Figure~\ref{fig4}, and the overall distribution is plotted in Figure~\ref{fig5}. Considering that the data used in the fitting basically cover the widest observation geometries achievable by a single station, and that the data reduction step is also conducive to homogenize the fitting weights on different observation geometries, the accuracy of our model is quite acceptable for the prediction of the brightnesses of satellites present in astronomical images.

\vspace{-6pt}  

\section{Photometric Distributions \label{Section 5}}

\subsection{Distributions in Body-Fixed Frame}

\begin{figure}[H]
\includegraphics[width=13.8 cm]{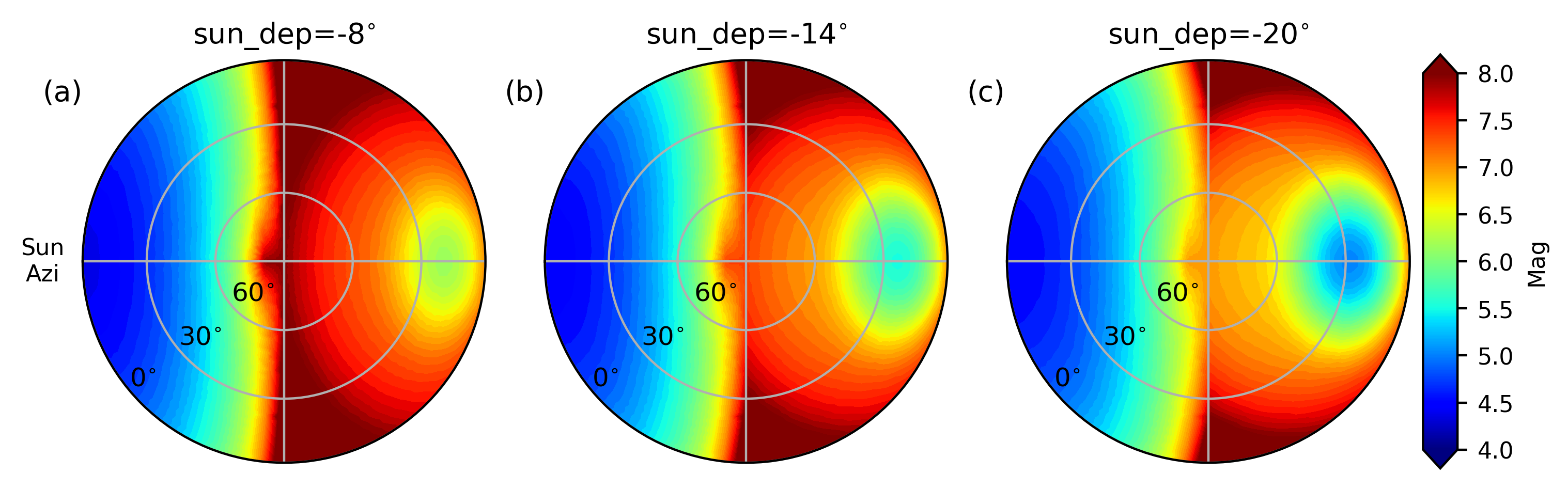}
\caption{{Distributions} 
 of satellite magnitude in the body-fixed frame at different solar altitudes. (\textbf{a}) The solar altitude is $-8^{\circ}$. (\textbf{b}) The solar altitude is $-14^{\circ}$. (\textbf{c}) The solar altitude is $-20^{\circ}$. The magnitude is corrected to a distance of 1,000 km. For satellites in 550 km orbits, the lowest solar altitude in the body-fixed frame accessible to ground-based observations is around $-20^{\circ}$, so the cases at lower altitudes are not plotted here.\label{fig6}}
\end{figure}

As the application of the model, the predicted photometric distributions are plotted at different solar altitudes in the satellite body-fixed coordinate system, which can intuitively display the contribution of the different satellite parts to whole brightness. As shown in Figure{~\ref{fig6}}, where the magnitude has been normalized to a 1000 km distance, the strong reflection from the front of the solar array makes the directions around the solar azimuth very bright. Since the size of the chassis is about one-sixth of the solar array, the anti-solar directions in which the back-scattering of the sail and the diffuse reflection of the chassis dominate, is usually darker in comparison, and only relatively brighter around the specular beam of the chassis. With the solar altitude decreasing, the reflection contribution from the chassis gradually increases and that of the sail is the opposite.

\subsection{Distributions in Observational Frame}

\begin{figure}[H]
\includegraphics[width=13.8 cm]{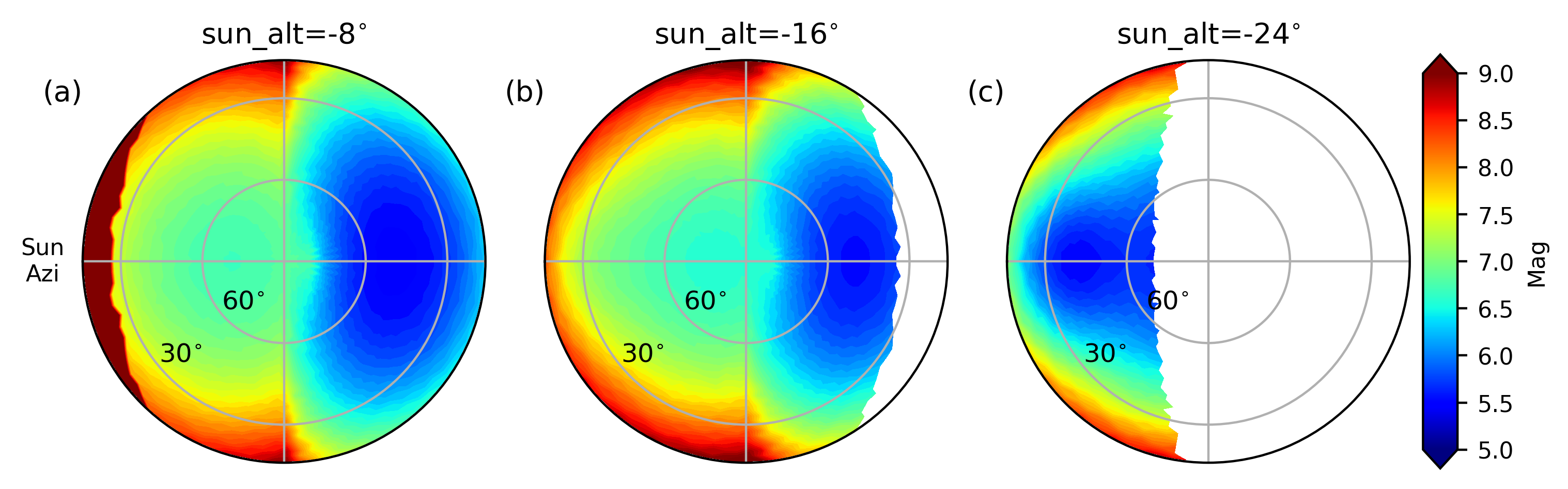}
\caption{{Distributions} 
 of the apparent magnitude of satellites in 550 km orbits at different solar altitudes in the Altaz frame. (\textbf{a}) The solar altitude is $-8^{\circ}$. (\textbf{b}) The solar altitude is $-16^{\circ}$. (\textbf{c}) The solar altitude is $-24^{\circ}$. At this altitude, the visible period of the satellite to the ground station extends from twilight to a solar altitude of around $-30^{\circ}$. The satellite apparent magnitude ranges from 5 mag to 9 mag.\label{fig7}}
\end{figure}

The apparent magnitudes of satellites on 550 km orbit over an observatory (where the satellites are visible) at different solar altitudes in Altaz frame are shown in Figure{~\ref{fig7}}. During twilight, satellites in the anti-solar direction are significantly brighter, and the reflection comes mainly from the front of the solar array. From twilight to nighttime, there is basically a gradual increase in brightness in the regions around solar azimuth, while the sky without illuminated satellites is also expanding in the anti-solar direction. These results are generally consistent with those in \cite{Cole}. If 7 mag is used as a criterion, the satellites are still overbright over a pretty wide range. 

\begin{figure}[H]
\includegraphics[width=13.8 cm]{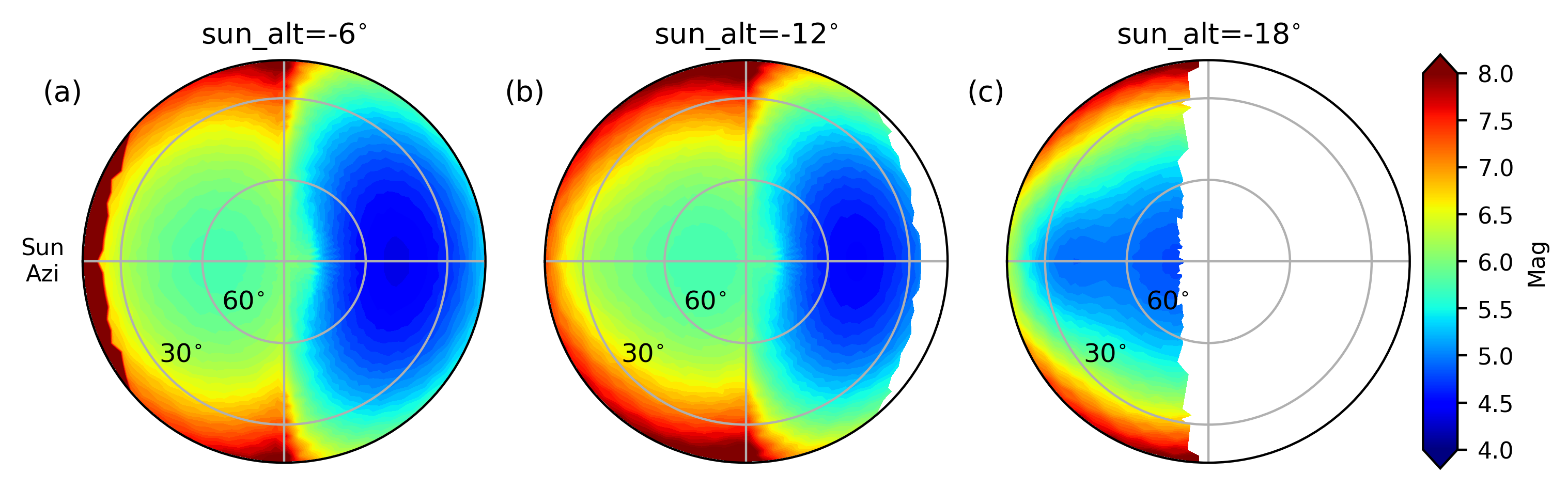}
\caption{{Distributions} 
 of the apparent magnitude of satellites in 340 km orbits at different solar altitudes in the Altaz frame. (\textbf{a}) The solar altitude is $-6^{\circ}$. (\textbf{b}) The solar altitude is $-12^{\circ}$. (\textbf{c}) The solar altitude is $-18^{\circ}$. At this altitude, the visible period of the satellite to the ground station extends from twilight to a solar altitude of around $-21^{\circ}$. The period of visibility is shorter than that of the 550 km satellites, and the apparent magnitude is relatively brighter within the range of 4 mag to 8 mag.\label{fig8}}
\end{figure}

\begin{figure}[H]
\includegraphics[width=13.8 cm]{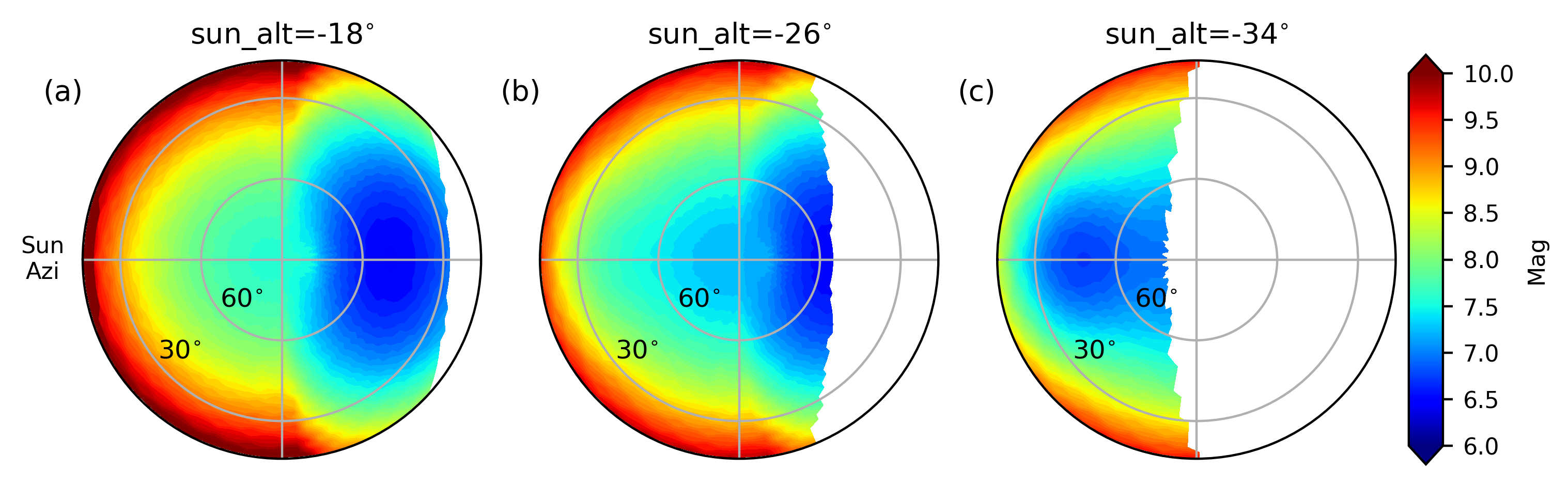}
\caption{{Distributions} 
 of the apparent magnitude of satellites in 1,200 km orbits at different solar altitudes in the Altaz frame. (\textbf{a}) The solar altitude is $-18^{\circ}$. (\textbf{b}) The solar altitude is $-26^{\circ}$. (\textbf{c}) The solar altitude is $-34^{\circ}$. At this altitude, the visible period of the satellite to the ground station extends from twilight to a solar altitude of around $-40^{\circ}$. The period of visibility is longer than that of the 550 km satellites, and the apparent magnitude is fainter within the range of 6 mag to 10 mag. However, the actual impact on astronomical observations needs to account for factors other than apparent magnitude, making the impact of satellites {in higher orbits more severe.}\label{fig9}}
\end{figure}

Starlink satellites are also planned to be launched into other lower or higher orbits, in which case the apparent magnitude distributions of the satellites, e.g., in the 340 km and 1200 km orbits, are shown in Figures{~\ref{fig8}} and{~\ref{fig9}}. It is clear that the higher the orbit, the further the satellite, and the lower the satellite magnitude, but meanwhile, the longer the time and the larger the sky area {will be} exposed in visible satellites. In fact, the differences in satellite apparent motion across the sky along with the out-of-focus effect make the actual impacts of satellite streaks on astronomical images dependent on more than just their apparent magnitude, as discussed in \cite{Tyson,Ragazzoni}. Therefore, satellites in higher orbits undoubtedly have more impact on astronomical observations because of their potentially brighter streak surface, as well as the explicitly greater temporal and spatial ranges affected.

\vspace{-9pt}  

\section{Discussion \label{Section 6}}

Through checking the O-C residuals, we found that the model is not well fit in some areas. Taking Figure{~\ref{fig4}} as an example, in the very narrow {regions} in the anti-solar direction at the lower observed altitude, the observed magnitudes are significantly smaller than those from the model. This region is within touching distance of the specular lobe of the chassis, while around this region, the situation rapidly changes to an observed magnitude larger than that modeled. This anomaly may be due to inaccurate modeling of the chassis specular reflection. Also, in the direction of solar azimuth plus or minus 90 degrees, our model using two thin panels as the satellite shape is almost incapable of producing reflections over there, but this region will still be somewhat bright in actual observations. It is notable that, in the neighborhood of solar azimuth of 70 degree altitude, the observed brightness exceeds significantly relative to the model, and it is difficult for our model to account for this {effect}, which may be due to the reflection from unknown satellite components that are not included in the model.

A {higher} fidelity BRDF model should come from direct measurements in laboratories on satellite models of a realistic shape and materials, which would obviously require the assistance of satellite manufacturers. Such a BRDF is presented in \cite{Fankhauser}, but the predictions of satellite magnitude are merely comparable to the models fit to the data. Possible reasons for this are not only the accuracy of the BRDF measurements themselves, but also other impact factors in the photometric model, such as the satellite attitude not being what is expected, insufficient modeling of other minor reflections, uncertainties in the observational data, etc. In short, until more accurate laboratory measurement models are made publicly available, the method of fitting models from massive observational data is absolutely indispensable and needs to be further investigated.

In this paper, only the data from the single MMT station are used to fit the model; although the observation geometry coverage is quite comprehensive, more data are still expected to be available to test the model. We will next collect other observations to verify the generalization capability of the model to unfamiliar data by carrying out new observations or collecting historical observations in the references, etc.

In addition, only V-band photometric data are used to build the model in this paper. For the reflectance properties of the satellite in other bands, multi-color photometric data are needed to build the model separately. This is undoubtedly a difficult task unless observations from different stations are coordinated, as recommended by \cite{dqs1,dqs2}. Until then, it seems feasible to use the solar spectrum to transform the satellite colors, provided that the satellite reflectance spectra remain in general agreement with the Sun \cite{Horiuchi1}, but this still needs to be confirmed by additional observations, i.e., by more satellite spectroscopic observations at different observation geometries.

\section{Conclusions \label{Section 7}}

The validation of brightness mitigation and simulations of satellite streaks in astronomical images depend on satellite photometric models, particularly those with high accuracy and spatial resolution. This paper fits models using millions of photometric observations with basic full geometric coverage for a single station. To enhance model accuracy and computational efficiency, we utilize data filtering and reduction procedures. Additionally, we account for the blocking effect of the chassis on the sail and the earthshine effect. The satellite attitude assumption, solar array opening angle assumption, sail specular reflection rationality, and earthshine necessity are also tested, and the results show that the satellite attitude is as expected, the sail is almost perpendicular to the chassis, and both sail specular reflection and chassis earthshine should be considered. The photometric distributions of the Starlink V1.5 satellite are plotted in both the body-fixed reference system and the observation reference system using the fitted model. The reflectance characteristics exhibited by the satellite are discussed based on these maps.

This paper has demonstrated the use of photometric models to characterize the distribution of the apparent magnitudes of satellites at different solar altitudes over an observatory. Photometric models also have potential applications in many other fields. For instance, modeling the photometry of different versions of satellites individually enables a global comparison of the effectiveness of brightness mitigation measures. This information will be valuable in designing targeted mitigation solutions in the future, as opposed to relying on histograms of brightness from limited observations. Photometric models are crucial in satellite streak simulations, as they allow astronomers to accurately evaluate the impact on their observations and develop effective streak-elimination techniques or avoidance strategies for bright satellites. In addition, space-based telescopes typically do not have such a large number of satellite observations to fully assess their impact \cite{Kruk}, and photometric models of satellites from ground-based observations will help to predict the brightness of satellite streaks encountered in space-based observations, especially the severe glints from specular reflections of the chassis. Finally, photometric models have potential applications in space situational awareness, e.g., anomalous photometric variations can be used to infer the operational status of satellites, and information in the models about the optical properties of satellite surface materials can be used to accurately model radiation pressures on satellites.

\vspace{6pt} 








\funding{{This research was supported by the Natural Science Foundation of Jiangsu Province of China (No. BK20221164) and the Pinghu Laboratory Open Foundation.}}

\dataavailability{The observations used in this paper are from the publicly available MMT database \url{http://mmt.favor2.info/satellites}, {accessed on 12 February 2024}. The TLE data used for the satellite position calculation are from space-track.org. The software used in this paper includes \textit{Numpy} \cite{numpy}, \textit{Scipy} \cite{scipy}, \textit{Astropy} \cite{astropy}, \textit{Skyfield} \cite{Skyfield}, \textit{Healpix} \cite{HEALPix}, \textit{Matplotlib} \cite{matplotlib}, and \textit{emcee} \cite{emcee}.}



\acknowledgments{We would like to thank Elena Katkova for the efforts in maintaining the MMT satellite photometric database. 
}

\conflictsofinterest{The authors declare no conflicts of interest.} 





\begin{adjustwidth}{-\extralength}{0cm}

\reftitle{References}

\PublishersNote{}
\end{adjustwidth}
\end{document}